# Saliency-based segmentation of dermoscopic images using color information

Giuliana Ramella

*National Research Council, Institute for the Application of Calculus, Italy;*
giuliana.ramella@cnr.it

Skin lesion segmentation is one of the crucial steps for an efficient non-invasive computer-aided early diagnosis of melanoma. This paper investigates how to use color information, besides saliency, for determining the pigmented lesion region automatically. Unlike most existing segmentation methods using only the saliency to discriminate against the skin lesion from the surrounding regions, we propose a novel method employing a binarization process coupled with new perceptual criteria, inspired by the human visual perception, related to the properties of saliency and color of the input image data distribution. As a means of refining the accuracy of the proposed method, the segmentation step is preceded by a pre-processing aimed at reducing the computation burden, removing artifacts, and improving contrast. We have assessed the method on two public databases, including 1497 dermoscopic images. We have also compared its performance with classical and recent saliency-based methods designed explicitly for dermoscopic images. The qualitative and quantitative evaluation indicates that the proposed method is promising since it produces an accurate skin lesion segmentation and performs satisfactorily compared to other existing saliency-based segmentation methods.

## 1. Introduction

Early diagnosis is a determining factor for melanoma prognosis since it allows treating this skin cancer at the first stage with a high cure rate through surgical removal of the tumor cells. The standard medical practice of melanoma diagnosis is the visual inspection of dermoscopic images by the skin doctor. This practice is a crucial part of the diagnosis process. However, besides being laborious and time-consuming, it has limited accuracy, is not easily playable, and requires adequate experience since it depends heavily on subjective judgment. Indeed, it has been proven that the dermoscopic diagnostic accuracy is lowered in the case of untrained dermatologists.

Usually, two different directions are taken to reduce the problems mentioned above. The first aims to improve diagnostic accuracy using a dermoscopic technique that displays morphological features not perceptible by the naked eye. Several studies [e.g., see Soyer et al. 2004; Perrinaud et al. 2007] have shown that the use of such techniques (i.e., solar scan, epiluminescence microscopy, and side transillumination) can improve the diagnostic accuracy of the visual examination by 20–30%, e.g., see [Binder et al. 1995; Menzies et al. 2005; Zouridakis et al. 2005]. The second focuses on improving the diagnostic assessment process and supporting dermatologists to perform a faster and more reliable skin analysis. For this purpose, new and innovative, fully automatic methods for segmentation, feature extraction, and classification of dermoscopic images have been proposed, see, e.g., [Premaladha et al. 2015; Mishra and Celebi 2016; Okur and Turkan 2018; Oliveira et al. 2018]. Most of these automatic methods also employ medical diagnostic techniques, for example, the 3-point or 7-point checklist, the Menzies' method, the ABCD rule, and the CASH algorithm [Soyer et al. 2004; Perrinaud et al. 2007; Scott Henning et al. 2007].

This paper follows the second direction and proposes, as the first stage to the computerized approach to melanoma diagnosis, a skin lesion segmentation method based on color and saliency information, named Saliency and Color Segmentation (SCS). This stage has a hit/effect on the performance of the successive steps, that is, feature extraction and classification.

Color image segmentation is a complex task, and extensive literature on this topic exists, e.g., see [Plataniotis and Venetsanopoulus 2013; Ramella and Sanniti di Baja 2013a; Ramella 2021b]. One of its applications is Automatic Skin Lesion Segmentation (ASLS). Although it has received considerable attention in the past decades and in recent literature, where one can find many segmentation methods based on different approaches, ASLS remains a challenging task. The main advantages and limits of ASLS methods have been discussed in many papers; see for instance [Silveira et al. 2009; Korotkov and Garcia 2012, Masood and Al-Jumaily 2013, Premaladha et al. 2015; Pereira et al. 2020]. A detailed review can be found in [Scott Henning et al. 2007; Silveira et al. 2009; Korotkov and Garcia 2012; Masood and Al-Jumaily 2013; Premaladha et al. 2015]. For a literature review, see Section 2 also, where a classification of ASLS methods based on the algorithmic approach is given. We remark that in this paper, our attention focuses on the existing class of saliency-based methods that detect skin lesions according to human visual perception since SCS is attributable to this class.

However, saliency is not the only feature helpful in handling skin lesion segmentation. Indeed, color information is one of the most cues humans use extensively to detect an object and, in general, for any image analysis task. For more information, as a sample, see [Arcelli et al. 2011, Plataniotis and Venetsanopoulus 2013; Ramella and Sanniti di Baja 2011, 2013a, 2013b, 2016b; Bruni et al. 2017; Ramella 2020] and the references therein. This applies especially in dermoscopy, where color information is used to detect skin lesions accurately by dermatologists [Soyer et al. 2004]. In this regard, the two following experimental results involving color information in dermoscopy have been considered in SCS. The first result, obtained by several experimental medical studies, concerns melanin: the most important chromophore in melanocytic neoplasm visible only by appropriate instrumentation. These studies revealed the melanin's color depends basically on its localization in the skin. The second refers to the role of color information in the diagnosis process. Color information is actually turned out to be the primary valuable feature for discriminating harmful and harmless lesions according to the aforementioned dermoscopic rules. Therefore, dermatologists indicate it as a key factor in skin lesion detection and evaluation; e.g., see [Soyer et al. 2004;

Ramella and Sanniti di Baja 2013a, 2016b; Bruni et al. 2017; Ramella 2020] and the references therein.

Notwithstanding, the existing saliency-based methods also using color information are mainly not designed explicitly for melanoma detection, e.g., see [Achanta et al. 2007; Harel et al. 2007; Zhang et al. 2017]. At the same time, the number of methods employing both saliency and color information tailored for skin lesion segmentation remains limited, e.g., see [Olugbara et al. 2018]. See Section 2 for more detail.

Starting from the above experimental evidence, following [Olugbara et al. 2018], we assume that, besides saliency, the role played by color information is crucial to segment skin lesions accurately. Thus, in this paper, we investigate how to employ saliency and color information for skin lesion segmentation. The rationale for employing both the saliency and color information is that this combined use can significantly improve the segmentation results, mainly when the lesion area has characterized by low contrast and/or different colors. Moreover, this enables to address the critical issues occurring in correspondence of salient regions sharing similar color features with the background.

Hence, we propose a method that, besides saliency, also looks at the color information, believed as extremely important, to detect the lesion region more precisely. To this purpose, we draw inspiration from the comprehensive literature regarding color image segmentation and human visual perception. Moreover, we consider the above experimental results to model the primary role of color information and define suitable criteria for the lesion detection process. Here, we extend and improve the preliminary version of the method proposed in [Ramella 2020], and, as mentioned above, we refer to this new version as SCS.

Similar to other existing methods for skin lesion segmentation, SCS consists of a pre-processing step aimed at preparing the image and a successive segmentation process to identify the skin lesion in terms of visual appearance (saliency) and color. The pre-processing step, including the saliency computation, is based on existing methods, while the segmentation step is wholly new and constitutes the core of the proposed method. See Section 3 for more details. We remark that we propose a new way to employ the saliency, calculating it by any valid existing method and then combining it with the color information of the detected salient region. To this aim, we resort to a set of perceptual criteria describing the lesion regions (see Section 3.2.2 for details).

The main innovative elements of the segmentation step are the following.
  a) The definition of the property of "color proximity," "peripheral component," and "kernel proximity" in terms of saliency and color information, inspired by human visual perception (see step B1.2).
  b) A set of perceptual criteria based on the above properties to mainly manage the case where the foreground/background transition of the skin lesion is not sharp and/or characterized by different colors (see steps B1.2 and B.1.3) or by colors very similar to those of background.
  c) The adopted algorithmic scheme (see Figure 2), based on the iterative delineation of the initial salient regions (see step B1.1) and on an adequate strategy consisting of two successive steps respectively related to the component insertion/elimination (see step B1.2) and the foreground expansion (see step B1.3)

Note that the segmentation process is flexible since one can enter multiple criteria and consider other types of information besides saliency and color. Moreover, the segmentation step is independent of how the saliency and other preliminary computations (resizing, color reduction, etc.) are performed in the pre-processing step. To give evidence to this latter aspect, we have adopted different computation methods [Dekker 1994; Harel et al. 2007; Achanta et al. 2009; Bruni et al. 2015; Ramella and Sanniti di Baja 2016a; Bruni et al. 2017] with various parameterizations in the pre-processing step, and we have evaluated the corresponding effects experimentally.

SCS has been tested on two publicly available databases $PH^2$ [Mendonca et al. 2015] and ISIC2016 [ISIC 2016], usually used in dermoscopic image processing. The results are evaluated in qualitative and quantitative terms also by comparing with other competitive methods. The experimental results confirm that:

1. SCS is an effective method and achieves good quantitative results with an adequate balance.
2. The employment of saliency and color information is helpful for skin lesion segmentation.
3. SCS is not influenced by the choice of the methods employed in the pre-processing step.
4. SCS has a competitive and satisfactory performance concerning other existing saliency-based methods.
5. SCS implementation is simple and quite fast since it does not require considerable computational power based on many parameters and labeled training images.

The remaining paper is organized as follows. Section 2 contains a literature review. Section 3 introduces the need for the pre-analysis phase (subsection 3.1), outlines the method SCS (subsection 3.2), and depicts the main results and methodological details (subsection 3.3). Subsection 3.2 is constituted by the subsections 3.2.1 and 3.2.2 that respectively describe the pre-processing step and the segmentation step, detailing the procedure and the adopted perceptual criteria also by examples. Section 4 specifies the adopted databases (subsection 4.1) and provides a quantitative (subsection 4.2) and qualitative (subsection 4.3) evaluation of experimental results, highlighting the pros and cons. Finally, Section 5 draws the final discussion and conclusions.

**2. Literature review**

We recall that the ASLS methods can be broadly classified based on their underlying algorithmic approach. The main categories are edge-based, thresholding, clustering, active contour models, supervised learning, and region-based. Frequently, methods belonging to distinct categories are combined in order to obtain maximum accuracy [Silveira et al. 2009].

The edge-based methods search for discontinuities in the intensity of the image pixels and their neighbors, usually adopting the magnitude of the gradient used to detect the edge of the lesion that constitutes the Region Of Interest (ROI); see, e.g. [Barcelos et al. 2009, Ganzeli et al. 2011]. In [Barcelos et al. 2009], the segmentation strategy is divided into two stages. The first detects the edges by applying the nonlinear diffusion model to selectively smooth the image and remove low-level contrast information, usually related to noises and hairs. The second stage finds out the lesion edges by applying the Canny edge detector to the previously smoothed image. The method proposed in [Yasmin et al.2011] converts the RGB image to a binary image by a thresholding technique, finds the xor image, and then finds the edges in this image using LOG detector. This category of edge-based methods is often combined with other techniques since they present identification problems on the skin lesion border and are too sensitive to noise.

The thresholding techniques are based on the input image's histogram and entail selecting a threshold value to separate the ROI in the input images; see, e.g. [Cavalcanti and Scharcanski 2013, Zortea et al. 2017]. In [Cavalcanti and Scharcanski 2013] preliminarily, a new image representation (channel) is created where the lesion features are more evident. Then, this new channel is thresholded, and the lesion border pre-detection is refined using a contour-detection algorithm followed by a process based on morphological operations. In [Zortea et al. 2017], skin segmentation is obtained by adapting Otsu's thresholding method and combining independent threshold estimates computed from histograms of different parts of a new intensity image suitably designed to discriminate lesions from background skin. Finally, a post-processing step that includes morphological filtering and a weighted scheme to select the most salient object is performed. Usually, the thresholding techniques detect lesions with very irregular borders with smaller size respect than they are and require successive elaborative steps to improve the shape of the segmented skin lesion.

In the clustering methods, each pixel is initially assigned to a cluster according to a similarity measure in terms of color information. Successively, a final segmented image is obtained using filtering and/or thresholding operations; see, e.g. [Agarwal et al. 2017, Zhou et al. 2013]. The method presented in [Agarwal et al. 2017] proposes a skin lesion segmentation method using the k-means clustering technique. Then, smoothing filtering and area thresholding are accomplished to reject the noisy pixels from the segmented image. In [Zhou et al. 2013], a mean shift-based segmentation algorithm is employed to locate the correct borders balancing the various forces resulting from the different energy terms when the contour reaches the equilibrium. Successively, a smoothness constraint process is carried out so to reduce over/under-segmentation effects. The methods belonging to the clustering category are also commonly employed in applications of a different type. The clustering methods are quite successful for the automated detection of the pigmented lesion if these are based on color besides spatial information.

In the class of methods based on active contours, the initial curve moves towards the object's boundaries through a suitable deformation, according to the adopted parametric/geometric model; see, e.g. [Oliveira et al. 2016, Ma and Tavares 2016]. In [Oliveira et al. 2016], an anisotropic diffusion filter reduces the noise present in the image under study. Then, the active contour model without edges is applied to the pre-processed image to segment the skin lesion. In [Ma and Tavares 2016], the RGB color space is converted to effectively use the color information to differentiate normal skin and healthy lesions. Afterward, the differences in the color channels are combined to define the speed function and the stopping criterion of the deformable model. For the methods based on active models, the difficulties are related to topological changes and large curvatures (parametric model) or high computational complexity (geometric models).

The learning approach is essentially based on convolutional neural networks (CNNs) and/or Fully Convolutional Networks (FCNs); see, e.g. [Yu et al. 2017, Kaymak et al. 2020]. Usually, the CNN is optimized by finding the global minimum (maximum) solution for the considered problem. Among the several types of optimization methods, the meta-heuristic ones seem to be the most promising since they are inspired by physics, nature, and human social behaviors. Examples of meta-heuristics optimization methods are particle swarm optimization [Tan et al. 2018], gray wolf optimization [Parsian et al. 2017], and world cup optimization algorithm [Razmjooy et al. 2018]. Besides melanoma detection, the advantages of these methods are proved in different applications, like optimizing the design for Energy Systems, Electric Vehicles, Power Transmission Systems, and dc/dc converters [Liu et al. 2020, Ye et al. 2020, Dehghani et al. 2021, Yang et al. 2021, Mehrpooya et al. 2021]. As concern skin lesion segmentation, as examples of CNN optimization, we mention the methods proposed in [Razmjooy et al. 2018, Xu et al. 2020]. In [Razmjooy et al. 2018], an optimized method based on soft computing is proposed. Specifically, after a pre-processing step for eliminating artifacts and noise, the world cup optimization algorithm is employed for optimizing the weights and the biases of the proposed network. Then, a post-processing step for the elimination of the extra areas is applied. In [Xu et al. 2020], an image segmentation step based on the convolutional neural network optimized by Satin Bowerbird Optimization (SBO) is applied after the image noise reduction step based on a median filter. The method is inspired by making the nests by the male bowerbird to attract the females. The learning methods reach good results, although in general require extensive learning based on many parameters and labeled training images.

The region-based methods group similar neighboring pixels into larger regions according to a given criterion; see, e.g. the classical methods [Iyatomi et al. 2008, Celebi et al. 2008]. The method proposed in [Iyatomi et al. 2008] consists of four phases: 1. initial tumor area detection by two filtering operations before the selection of a threshold; 2. regionalization by merging small isolated regions created in the first phase; 3. tumor area selection by selecting appropriate areas according to predefined criteria; (4) region growing. In [Celebi et al. 2008], after a pre-processing step including the smoothing of the image, a fast and unsupervised technique is applied based on the statistical region merging. Next, to eliminate spurious detected regions belonging to the background, a post-processing step is executed. Typically, the region-based methods distinguish the lesion components (regions) by standard image processing techniques such as, for instance, statistical region merging [Celebi et al. 2008], modified JSEG [Celebi et al. 2007], watershed [Wang et al. 2011], and complex networks [Aksac et al. 2017]. Thus, the identification problems of the region-based methods are directly connected to the kind of employed processing technique.

Interesting results have been obtained by region-based methods that use the saliency to identify relevant regions as those that are visually more distinctive due to their contrast; see, e.g., [Yang 2013; Zhu et al. 2014; Cheng et al. 2015; Ahn et al. 2017; Fan et al. 2017; Hu et al. 2019]. These methods have become a prominent tool for medical image analysis since saliency focuses on image regions that carry helpful information according to human visual perception. In this paper, our attention focuses on this latter class of methods, usually indicated simply as saliency-based methods. Specifically, in this subsection, we describe shortly the methods considered in the validation phase of SCS, namely MR [Yang et al. 2013], RDB [Zhu et al. 2014], RC [Cheng et al. 2015], RSSLS [Ahn et al. 2017], Fan [Fan et al. 2017] and Hu [Hu et al. 2019].

Classical saliency-based methods belonging to the region-based method class (e.g., MR, RDB, RC) have broad application in image segmentation. In MR, relevance to the given seeds or queries is employed to define the saliency of the image elements (pixels or region). Then, the similarity of image elements with foreground/background cues via the graph-based manifold is ranked, and a close-loop graph with superpixels as nodes is employed to represent the image. Finally, the nodes are ranked by affinity matrices according to the similarity to the background and the foreground queries. In RDB, a measure to estimate how heavily a region is connected to the image borders is adopted. Then, a principled optimization framework to incorporate multiple low-level cues, including the background measure, is employed to obtain clean and uniform saliency maps. The RC method is based on the computation of global contrast difference and spatially weighted coherence scores. Then, the salient regions' detection is obtained using the saliency maps to initialize a new iterative version of GrabCut [Rother 2004].

Applying saliency-based segmentation methods (e.g., RSSLS, Fan, Hu) to dermoscopy is relatively new. The methods in this class compute the salient region in an image according to human visual perception such that a salient (not salient) part becomes a foreground (background) region, namely skin lesion (healthy lesion). In RSSLS, the saliency is computed using the reconstruction errors derived from a sparse representation model and a novel background detection process. The shape and the borders of the lesion are delineated in a Bayesian framework. In Fan, saliency combined with the Otsu threshold is proposed for the automatic skin lesion segmentation. The method includes two phases: enhancement and segmentation. In the first phase (enhancement), the color and brightness saliency maps are computed and fused to obtain the enhanced image. In the second phase (segmentation), to extract more accurate lesion borders, an optimization function is designed to adjust the traditional Otsu threshold method according to the histogram distribution of the enhanced image. In Hu, the contrast of the skin lesion and healthy skin is increased by fusing the corresponding saliency maps. Then, an adaptive wavelet-based thresholding method is employed.

Notwithstanding, color information is not usually employed in saliency-based methods explicitly designed for lesion segmentation. In general, in most existing methods belonging to this category, the perceptual saliency is computed by localizing the high contrast and brightness region between the background and the foreground of the skin lesion, disregarding the color information. On the other hand, color information is widely used in medical image analysis and, specifically, in skin lesion segmentation in not saliency-based methods, e.g., see [Itti et al. 1998, Garnavi et al. 2011, Zortea et al. 2011]. Moreover, this information is also used in classical saliency-based methods not applied to dermoscopic images, e.g., see [Achanta et al. 2007, Harel et al. 2007, Zhang et al. 2017]. As an example of an exception to this general trend, we point out that a perceptual color difference saliency with morphological analysis for dermoscopic images has been recently proposed in [Olugbara et al. 2018].

## 3. The proposed method

### 3.1. Pre-analysis

The color image segmentation problem becomes hard to manage when applied to dermoscopic images due to the frequently "inconsistent" imaging conditions [Celebi and Schaefer 2013]. In particular, the images can be acquired under different illumination conditions and may have poor resolution. An additional complex problem is related to artifacts, such as bubbles, hair, shadows, reflections, which may negatively affect and disrupt the analysis of the skin lesions.

Some artifacts, e.g., dark corners, ink markers, rulers, and bubbles, are caused directly by the imaging technique. Specifically, the presence of hair (often also considered as an artifact) and/or different colors inside the lesion are due to the nature of the image, while the low contrast at the border separating the lesion from the surrounding healthy skin is usually due to different conditions of illumination, contrast, and noise. In Figure 1, some examples are shown.

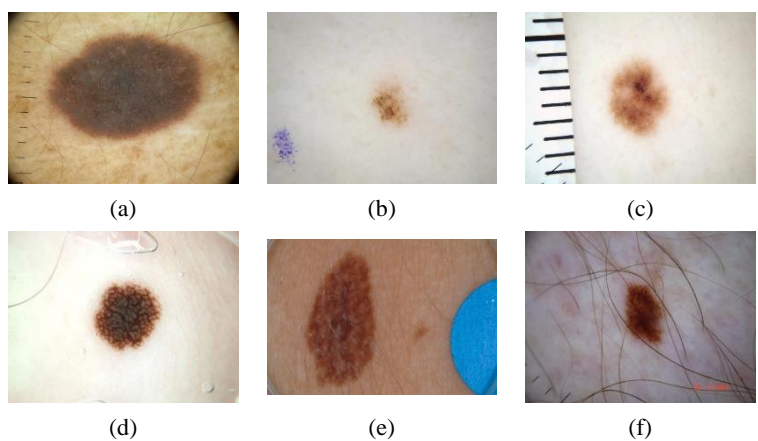

**Figure 1.** Examples of possible artifacts and aberrations.

Since artifacts and the mentioned aberrations dramatically affect the accuracy of the segmentation, most skin lesion segmentation methods are designed so that all artifacts and undesired effects are removed, or at least mitigated, before applying the segmentation process. Consequently, almost all dermoscopic image segmentation methods include a pre-processing step to improve the quality of the image and remove the effects of artifacts/aberrations.

This step is typically based on standard image analysis techniques such as median filtering [Fan et al. 2017], color correction [Quintana et al. 2011], illumination correction [Glaister et al. 2013], contrast enhancement [Barata et al. 2015], and hair removal [Lee et al. 1997, Ramella 2021a]. Any residual effects should then be suitably considered and adequately treated during the segmentation phase to disrupt as little as possible the skin lesion detection. Moreover, a post-processing step, primarily based on region merging and/or morphological operations, is often applied to improve skin lesion region detection.

*3.2. Structure of the proposed method*

As mentioned above, SCS aims to segment dark skin lesions in dermoscopic images using two different types of information (saliency and color). It follows the algorithmic scheme described in Section 2.1; namely, it consists of two consecutive main steps: pre-processing and segmentation. The pre-processing step (see Section 2.2.1) aims to prepare the image by eliminating artifacts (hair included) and computing the Saliency Map (SM). The core step is the successive segmentation (see Section 2.2.2) which determines, in terms of visual appearance (saliency) and color, a Binary Mask (BM) representing the localization of the skin lesion, which is then employed to determine the skin lesion. This innovative step allows overcoming, in most cases, the problems due to different colors, low contrast, and lesion color similarity with the background. The main structure of SCS is presented in Figure 2. This figure shows that the input image is entered into the pre-processing step and then into the segmentation step.

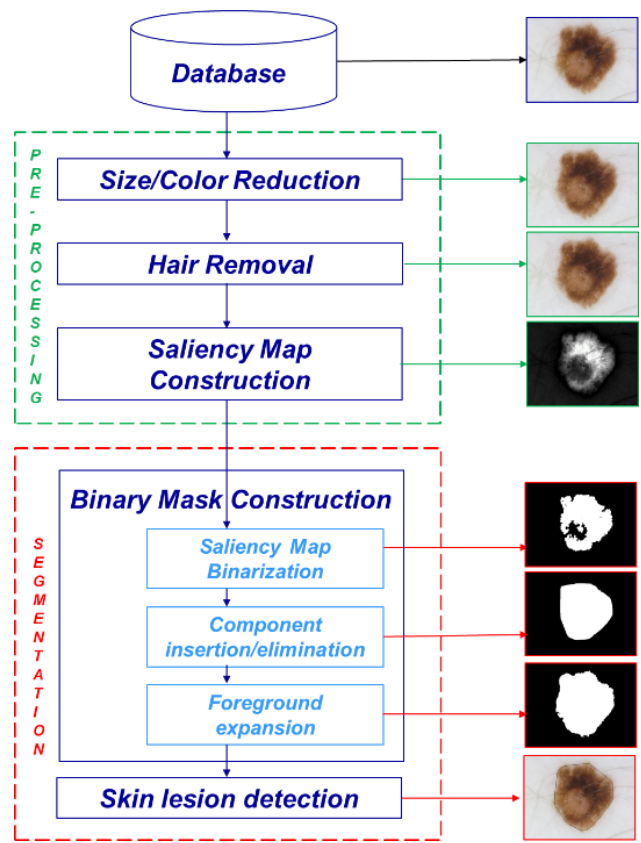

**Figure 2.** Flowchart of the proposed method, SCS**.**

The steps of this flowchart can be summarized as:

***Pre-processing step*** (see Section 3.2.1)
Step A.1: Size/Color Reduction
Step A.2: Hair Removal
Step A.3: Saliency Map Construction

***Segmentation step*** (see Section 3.2.2)
Step B.1: Binary Map Construction
Step B.2: Skin lesion Detection

where step B.1 is structured in the following substeps:

Sub-step B.1.1: Saliency Map Binarization
Sub-step B.1.2: Component insertion/elimination
Sub-step B.1.3: Foreground expansion

In Step B.1.2, we remark that we introduce i) a criterion, named *Color proximity criterion* (1) based on the *Color proximity property*; ii) a criterion, named *Peripheral component criterion* (2), based on the *Peripheral/kernel component definition* and the *Kernel proximity property*. See below for more details. Also, in B.1.3, we employ the property of color proximity to apply the criterion (3) and then (4).

Note that in the following subsections of the current subsection 3.2, we describe the main methodological aspects of each step and we show some results.

### 3.2.1. Pre-processing step

The pre-processing step consists of three sub-steps. The first sub-step, namely Size/Color Reduction (A.1), limits the successive steps' computation burden by reducing the size and number of colors. This early sub-step is an optional but highly recommended operation since it significantly bounds also the computation time. The target of the second sub-step, namely Hair Removal (A.2), is to remove or, at least to limit, the sign of hair since the resulting segmentation could be affected by it. Finally, the third sub-step, namely Saliency Map Construction (A.3), is for the saliency computation, one of the two primary information considered helpful for skin lesion detection. In more detail, the pre-processing step can be outlined as follows.

Step A.1 - Size/Color Reduction. The image size is reduced so that the maximum between the number of rows and the number of columns is equal to a value fixed by the user, say *Maxdim*. Downsampling is done using bicubic interpolation. The reduction of the number of colors, namely *colnum*, specified by the user, is obtained by a Color Quantization (CQ) method. Four different CQ methods are available for this purpose: [Dekker 1994; Ramella and Sanniti di Baja 2016a; Bruni et al. 2015, 2017].

Step A.2 - Hair Removal. The hair present in the lesion image is removed, and the position of the image in correspondence with the removed hair is restored by the algorithm proposed in [Lee et al. 1997]. Note that this step is necessary since the presence of hair, which typically appears superimposed and/or close to the lesion, can lead to not suitable segmentation, e.g., see the example shown in Figure 1f). Unfortunately, hair removal is not always wholly successful. The influence of hair not entirely removed and other artifacts is strongly reduced in the following steps by suitable operations based on saliency and color information.

Step A.3 - Saliency Map Construction. SM with well-defined boundaries of salient objects is computed by the method proposed in [Harel et al. 2007] or [Achanta et al. 2009]. Then, the Saliency Map is enhanced by increasing the contrast in the following way: the values of the input intensity image are mapped to new values obtained by saturating the bottom 1% and the top 1% of all pixel values.

*3.2.2 Segmentation step*

The segmentation step consists of two main sub-steps: Binary Mask Construction (B.1) and Skin Lesion Detection (B.2). The first sub-step (B.1) detects the Binary Mask delimiting the region of interest and then determines the skin lesion (B.2).

Step B.1 - Binary Mask Construction. Three main sub-steps are required to compute the Binary Mask: Saliency Map Binarization (B.1.1), Component inclusion/elimination (B.1.2), Foreground Expansion (B.1.3). The first sub-step (B1.1) extracts the more salient regions and detects the initial Binary Mask. The second (B.1.2) and the third sub-steps (B.1.3) are devoted respectively: i) to eliminate false salient regions due to inadequate illumination or due to low contrast; ii) to insert/eliminate regions in the binary mask by considering saliency and color information. In more detail, the step of Binary Mask Construction can be described as follows.

Sub-step B.1.1 - Saliency Map Binarization. The strategy initially adopted in this step is to identify the "core" of the lesion, the initial Binary Mask, by an iterated process based on saliency information. Thus, SM undergoes an iterated binarization process aimed at extracting the more salient regions. At each iteration, SM is binarized in the following way. All pixels with saliency value greater than the average saliency value $\mu_s$ are recorded into a Binary Mask (BM), initially empty, and considered belonging to the foreground of BM, shortly indicated as BM-foreground (BM$f$). All other pixels are assigned to the background of BM, shortly indicated as BM-background (BM$b$).

Indicated by $R_f$ ($R_b$) the connected components of BM$_f$(BM$_b$), the average saliency value $\mu_s$ is newly computed by ignoring the pixels belonging to connected components $R_f$ which include pixels on the image frame. Generally, a smaller average saliency value $\mu_s$ is obtained with the result that, when the SM is newly binarized for the computed new saliency value, a larger number of pixels overcome this new threshold value. Therefore, these pixels are considered as potentially belonging to BM$f$. The process for saliency map binarization is iterated as far as connected components, including pixels of the image frame, are detected. Thus, BM constitutes the initial Binary Mask.

Note that this iterated process is necessary to manage some images where the average saliency value is strongly conditioned by the presence of artifacts caused directly by the adopted image acquisition technique. For example, it happens when dark corners are present (see Figure 1a) or, even worse, when the region of interest for saliency map thresholding constitutes a circular portion of an otherwise entirely dark image. In correspondence to these dark artifacts having a relatively high saliency value, the average saliency value would be higher than the average saliency value characterizing the pixels in the skin lesion. Accordingly, the number of pixels that would be assigned to $BM_f$ is remarkably smaller than expected. For this reason, saliency map binarization involves a process aimed at identifying a Binary Mask including all pixels whose saliency value should not be counted for average saliency value computation. Moreover, an iterated process is performed to identify BM since not all pixels of the artifacts are characterized by the same (high) saliency value.

Successively, according to the assumption regarding the combined use of saliency and color information (see the introduction and Section 2), we introduce some criteria inspired by human visual perception and related to both the types of information to identify the skin lesion. Hence, in successive steps (*B.1.2* and *B1.3*), connected components (regions) are eventually aggregated to BM or eliminated based on saliency/color information and the visual contest.

Sub-step B.1.2 - Component insertion/elimination. A critical configuration for the construction of BM can occur when the $BM_b$, usually lighter than the $BM_f$, includes some lighter regions than the remaining background regions. Indeed, since salient regions are visually more evident than the surrounding areas, both pixels inside the lesion (hence characterized by darker colors) and pixels in the lightest portions of the BMb may have high enough saliency to make them potentially assignable to BM$f$ during binarization. This problem is solved by introducing *the property of color proximity* and by resorting to the *color proximity criterion* (1) defined as follows.

*Color proximity property* - The lesion can also include regions with not high saliency, but it can have average color "perceptually close" to the color of the skin lesion already detected.

More specifically, denoted by:

C* - the darkest detected color of $BM_f$,
$C_f$ - the average color associated with each $R_f$ of $BM_f$, for $f=1,…,N$
$d_f$ ($C_f$, C*) - the distance of the average color $C_f$ from C*, for $f=1,…,N$
dmin= *min*{ $d_f$ ($C_f$, C*)}, for $f=1,…,N$
dmax= *max*{ $d_f$ ($C_f$, C*)}, for $f=1,…,N$
Δ = dmax – dmin
δ = *mean*(dmax,dmin)

we postulate that "a small value of Δ implies that the BM$f$ includes connected components characterized by average colors perceptually close that should be maintained in the BM$f$, while a large Δ implies that rather different average colors characterize the components". Hence, some of these components should be disregarded and not considered as belonging to the lesion.

According to this *color-proximity property*, we adopt the following *color proximity criterion.*

*Color proximity criterion* - If Δ is greater than a threshold value $T_c$, remove from $BM_f$ the components $R_f$ characterized by an average color $C_f$ larger than δ:

$$\text{if } \Delta > T_c \ \& \ (\exists \ R_f \ \epsilon \ BM_f \ : \ C_f > \delta) \Rightarrow (R_f \notin BM_f) \ \& \ (R_f \in BM_b) \quad (1)$$

where $T_c$ is a prefixed threshold value on the color difference perceptible to the human eye.

As a result, we accept as belonging to $BM_f$ also some pixels characterized by low saliency values that prevented their extraction during binarization, provided that their colors are closer to the average color of $BM_f$ than the average color of the $BM_b$.

Currently, BM is likely to consist of some components, all characterized by saliency or color compatible with those expected for the skin lesion to be segmented. However, not all these components belong to the skin lesion. Some peripheral parts can be noisy regions characterized by saliency or color similar to those in the skin lesion. To keep only the relevant components, we distinguish the remaining $R_f$ in the kernel and peripheral components according to the following definition.

*Peripheral/kernel component definition* - A component $R_f$ including less of $T_n$ pixels with saliency greater than $2\mu_s$ is considered as peripheral component" (i.e., belonging to the set PC of the peripheral components of $BM_f$); otherwise, it is a kernel component (i.e., belonging to the set KC of the kernel components of $BM_f$).

Moreover, denoted by:

CH(KC, $R_f$) - the convex hull of KC plus the peripheral component $R_f$
CH(KC) - the convex hull of the kernel components KC
A[KC] - the total area of the kernel components KC
A[$R_f$] - the area of the $R_f$ component
A[CH(KC, $R_f$)] - the area of CH(KC, $R_f$)
A[CH(CK)] - the area of CH(CK)
AD= A[CH(KC, $R_f$)]- A[CH(KC)]- A[$R_f$]
rig - the number of rows of the input image
col - the number of columns of the input image
minsize=*min*(rig,col);

we estimate the proximity of a peripheral component $R_f$ to the kernel KC according to the following *kernel proximity property*.

*Kernel proximity property* - $R_f$ is sufficiently close to KC if A[$R_f$] is less than AD, i.e., AD is sufficiently small".

Then, to eventually ascribe the peripheral components to $BM_b$, based on the *peripheral/kernel component* definition and the *proximity kernel proximity*, we apply the following criterion, namely the *peripheral component criterion* (2).

*Peripheral component criterion* - If a peripheral component $R_f$ has area A[$R_f$] less than of a significant percentage, namely $\theta_l$, of minsize or has area A[$R_f$] larger than A[KC] or $R_f$ is sufficiently close to KC, (i.e., A[$R_f$] <AD) then ascribe it to $BM_b$:

if (A[$R_f$] < $\theta_1$ minsize || A[$R_f$] >A[KC] || A[$R_f$] <AD) $\Rightarrow$ ($R_f \notin BM_f$) & ($R_f \in BM_b$)    (2)

An example is shown in Figure 3-4, step B.1.2.

<u>Substep B.1.3 - Foreground expansion.</u> In many cases, the segmented $BM_f$ obtained at the end of the previous step is already close enough to the desired result. However, there are skin lesions for which the color transition from foreground to background is not sharp. In these cases, we may note a kind of band, namely *Color Transition Region* (*$CTR_f$*), surrounding the detected foreground in SM. Thus, we also employ color information and the property of *color proximity* to detect lesion components in correspondence with the region surrounding the detected $BM_f$ having a not-sharp color transition from lesion foreground to background.

In particular, the surrounding region $CTR_f$ is detected by considering from saliency map SM pixels not belonging to $BM_f$, having low saliency values, say smaller than a prefixed value $T_s$, but having quite different colors according to the above property of *color proximity*. Therefore, denoted by:

$C_m$ - the average color of the current $CRT_f$
$C_b$ - the average color of $CRT_b$
d ($C_b$, $C_m$) - the distance of the average color $C_b$ from $C_m$
$C_p$ - the color of a pixel p $\in$ $CRT_f$
dp ($C_p$, $C_m$) - the distance of the average color $C_p$ from $C_m$

we filter out from $CRT_f$ the pixels that cannot be assigned to $CRT_f$, that is, any pixel p of $CRT_f$ whose color $C_p$ has distance from $C_m$ larger than a significant percentage, namely $\theta_2$, of d($C_b$, $C_m$) is assigned to $CRT_b$:

$\forall$ p $\in$ $CRT_f$ : $d_p$ ($C._b$, $C_m$)> $\theta_2$ . d ($C_b$, $C_m$) $\Rightarrow$ (p $\notin$ $CRT_f$) & (p $\in$ $CRT_b$)    (3)

Then, the remaining pixels of $CRT_f$ not belonging to $BM_f$ but connected to $BM_f$ are assigned to $BM_f$ according to (4)

$\forall$ p $\in$ $CRT_f$, p $\notin$ $BM_f$: p connected to $BM_f$ $\Rightarrow$ p $\in$ $BM_f$    (4)

As an example, see Figure 3-4, step B.1.3.

<u>Step B.2 - Skin Lesion Detection.</u> This step refines the obtained BM to select the lesion regions among the possible candidates, improve the final shape, and better comply with the region manually delineated by the skin specialist. To this purpose, morphological operations are applied to smooth the contour of the segmented skin lesion and fill possibly existing small holes.

If more than one component is obtained, since only one lesion component is expected for the considered databases (ISIC2016 and PH2), the one with the largest area is selected.

The convex hull of such a component is finally computed. Thus, it constitutes the result of segmentation. In the alternative, if the considered database foresees more than one possible lesion, this step is skipped.

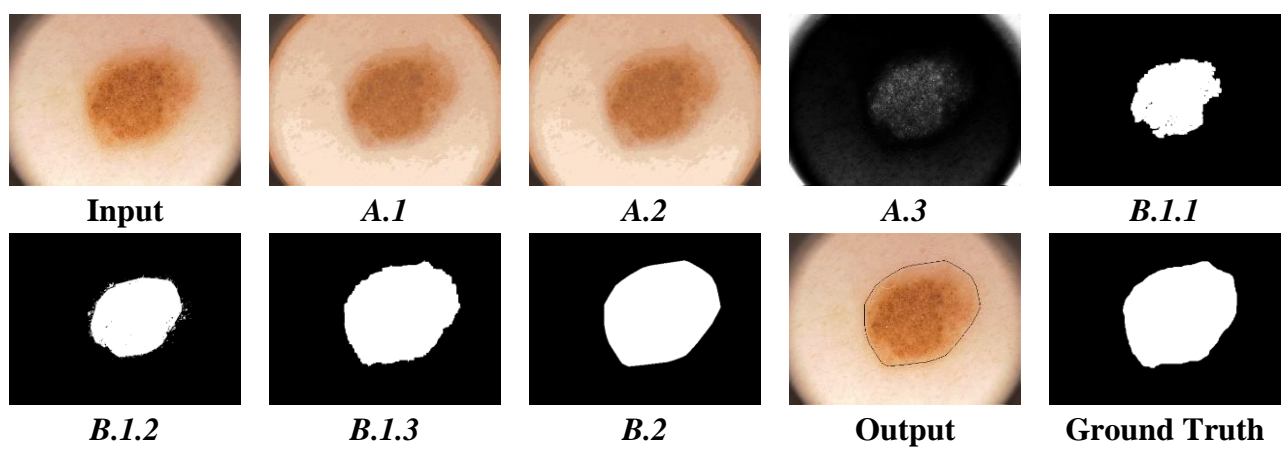

**Figure 3.** Results of each step by SCS on an image of PH$^2$ database and the corresponding ground truth.

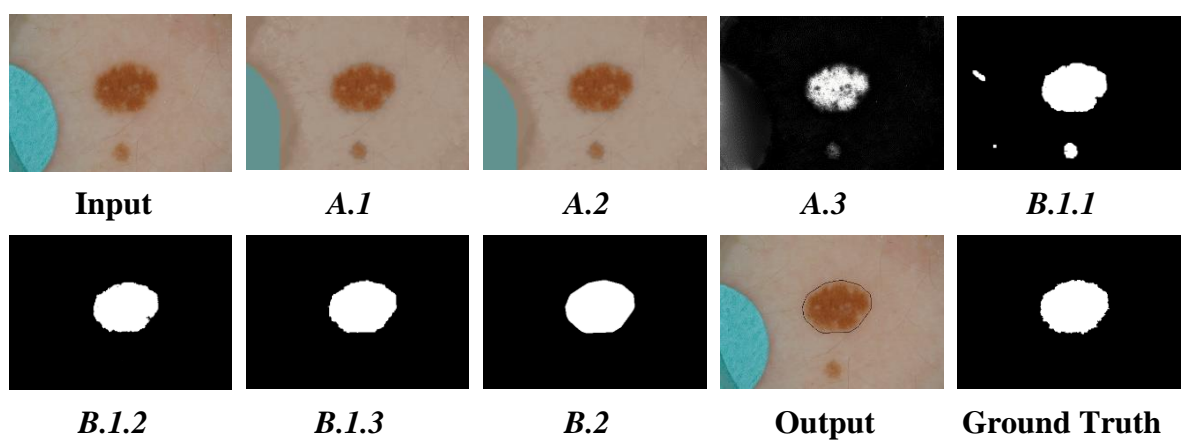

**Figure 4.** Results of each step by SCS on an image of ISIC2016 database and the corresponding ground truth.

### 3.3. Results and methodological details

In Figure 3 and Figure 4, the SCS performance results in each step are shown for an image of PH$^2$ and ISIC2016, respectively. In Figure 5, a few examples of the results under varying critical conditions (see Figure 1), such as the presence of a dark halo, blue ink stains, a ruler, air bubbles, colored disks, thick hair, are shown. Other results with different performances are shown in Figure 6. For each example in Figure 5-6, the quality performance evaluation metrics, mentioned and defined respectively in Section 4 and Table 1, are given. In Figure 6, a comparison of the segmentation result with the ground truth (see the ultimate column) is also visualized using the color white for True Positive (TP) pixels, red for False Positive (FP) pixels, and green for False Negative (FN) pixels. In Figure 3-6, the detected boundary of the segmented skin lesion is shown superimposed onto the input.

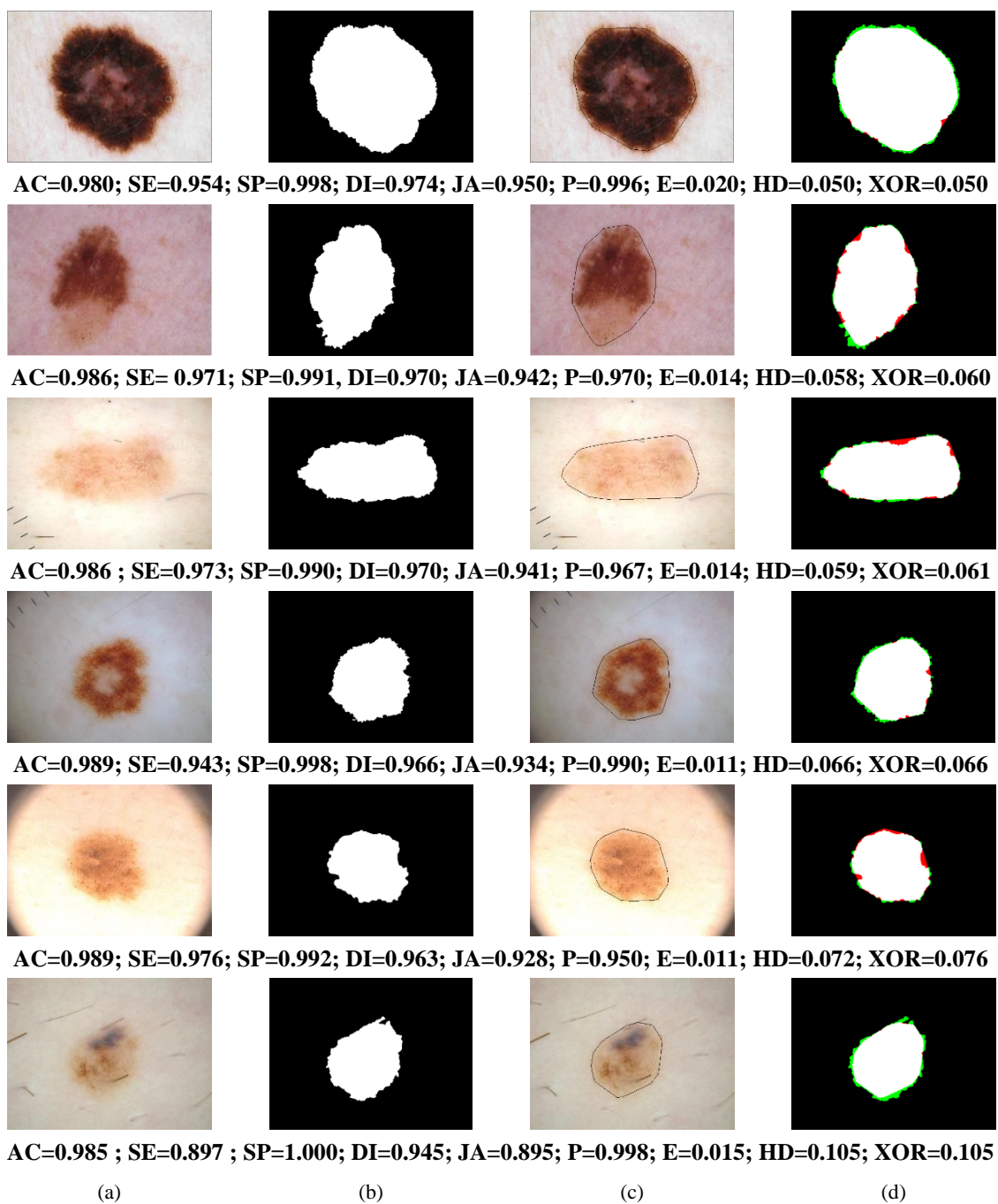

|   |   |   |   |
|---|---|---|---|
| (a) | (b) | (c) | (d) |

**Figure 5.** a) Inputs; b) ground truths; c) different performance of SCS on images of the two databases, where the detected boundary is shown superimposed onto the input; d) comparison of the segmentation results with the ground truth, where white is used for TP, red for FP and green for FN pixels.

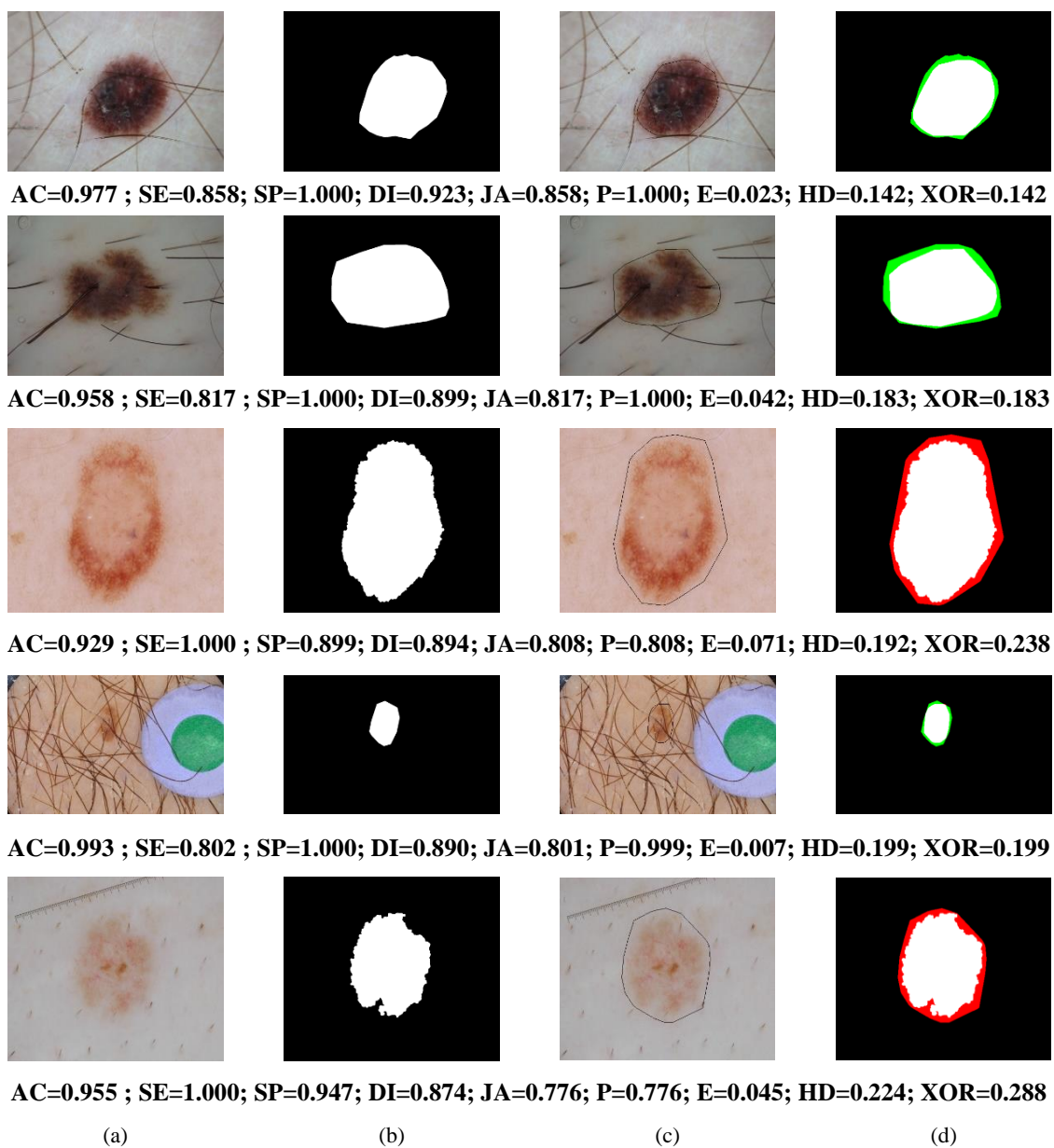

**AC=0.977 ; SE=0.858; SP=1.000; DI=0.923; JA=0.858; P=1.000; E=0.023; HD=0.142; XOR=0.142**

**AC=0.958 ; SE=0.817 ; SP=1.000; DI=0.899; JA=0.817; P=1.000; E=0.042; HD=0.183; XOR=0.183**

**AC=0.929 ; SE=1.000 ; SP=0.899; DI=0.894; JA=0.808; P=0.808; E=0.071; HD=0.192; XOR=0.238**

**AC=0.993 ; SE=0.802 ; SP=1.000; DI=0.890; JA=0.801; P=0.999; E=0.007; HD=0.199; XOR=0.199**

**AC=0.955 ; SE=1.000; SP=0.947; DI=0.874; JA=0.776; P=0.776; E=0.045; HD=0.224; XOR=0.288**

(a)  (b)  (c)  (d)

**Figure 5 (continued).**

We remark that:

a) The reduction of noise on the contour and the elimination of isolated and small regions in B.2 helps to improve skin lesion detection. On the contrary, if more than one lesion region has been detected in B.2, selecting the component with the largest area is necessary for the databases under consideration.

b) The convex hull is computed since, especially for ISIC2016 images, even when the skin contour is visually well delineated, in most cases, the outline provided by convex hull usually better adapts to the skin border of ground truth [Guarracino and Maddalena 2019]. See the examples shown in Figures 3-6.

c) The segmentation step requires to establish the values of five perceptive threshold values: $T_c$ (relative to the color difference perceptible to the human eye), $T_n$ (relative to the minimum number of pixels with high saliency of a peripheral component), $\theta_l$ (relative to the percent of the minimal area visible to the human eye), $T_s$ (relative to the saliency values considerable as not salient), θ$_2$ (relative to the percent on color distance perceptible to the human eye). The choice of the value of these parameters depends on the class of images to handle. Their values have been experimentally determined for the databases PH$^2$ and ISIC2016 (see subsection 3.2).

**Table 1.** Evaluation Metrics, where FP, FN, TP and TN respectively denote False Positive, False Negative, True Positive and True Negative assessments.

| Symbol | Metric | Definition |
|---|---|---|
| AC | Pixel-level accuracy | $AC = \dfrac{TP + TN}{TP + FP + TN + FN}$ |
| SE | Sensitivity | $SE = \dfrac{TP}{TP + FN}$ |
| SP | Specificity | $SP = \dfrac{TN}{TN + FP}$ |
| DI | Dice coefficient | $DI = \dfrac{2 * TP}{2 * TP + FN + FP}$ |
| JA | Jaccard index | $JA = \dfrac{TP}{TP + FN + FP}$ |
| P | Precision | $P = \dfrac{TP}{TP + FP}$ |
| E | Error | $E = \dfrac{FP + FN}{TP + FP + TN + FN}$ |
| HD | Hammoude distance | $HD = \dfrac{FP + FN}{TP + FN + FP}$ |
| XOR | XOR | $XOR = \dfrac{FP + FN}{TP + FN}$ |

## 4. Performance evaluation

### 4.1. Databases

We have tested our method on two publicly available databases of dermoscopic images: PH$^2$ [Mendonca et al. 2015] and ISIC2016 [ISIC 2016].

PH$^2$ is a dermoscopic image database acquired at the Dermatology Service of Hospital Pedro Hispano to support comparative studies on segmentation/classification methods. This database includes clinical/histological diagnosis, medical annotation, and the evaluation of many dermoscopic criteria. It provides 200 dermoscopic RGB images and the corresponding ground truth, including 80 atypical nevi, 80 common nevi, and 40 melanomas. All images are 8-bit RGB and have resolution 760x560 pixels.

ISIC2016 is one of the largest databases of dermoscopic images of skin lesions with quality-controlled held by the International Symposium on Biomedical Imaging (ISBI) to improve melanoma diagnosis. It includes images representative of both benign and malignant skin lesions. Moreover, the corresponding ground truth of each image is also available. Parts of the ISIC2016 dataset used in this paper consist of 397 (75 melanomas) and 900 (173 melanomas) annotated images for testing and training, respectively. The images are 8-bit RGB and have a size ranging from 542x718 to 2848x4288.

PH$^2$ and ISIC2016 datasets contain numerous images with complex backgrounds, complicated skin conditions, and various artifacts and aberrations. We point out that different high levels of detail characterize the ground truth images of these datasets. Indeed, mainly in ISIC2016, the lesion's boundary of some ground truth images is delineated accurately (typical of segmentation obtained by an automatic segmentation software). In contrast, the border of other ground truth images is more roughly approximated (distinctive of segmentation manually achieved by dermatologists).

### 4.2. Qualitative evaluation

The segmentation results obtained by SCS have been qualitatively analyzed and compared with the segmentation results of the existing mentioned saliency-based segmentation methods (MR, RDB, RC, Fan, RSSLS, Hu) using the images available in PH$^2$ and ISIC2016 databases. For most of the examined images, as you can also see from the examples shown in Figures 3-6, the skin lesions are correctly detected by SCS with well-defined boundaries. Furthermore, SCS is able to distinguish the skin lesion in the images, including hair and/or characterized by similar colors in the lesion region and background. Finally, SCS achieves adequate results for images having high contrast, e.g., see Figure 6.

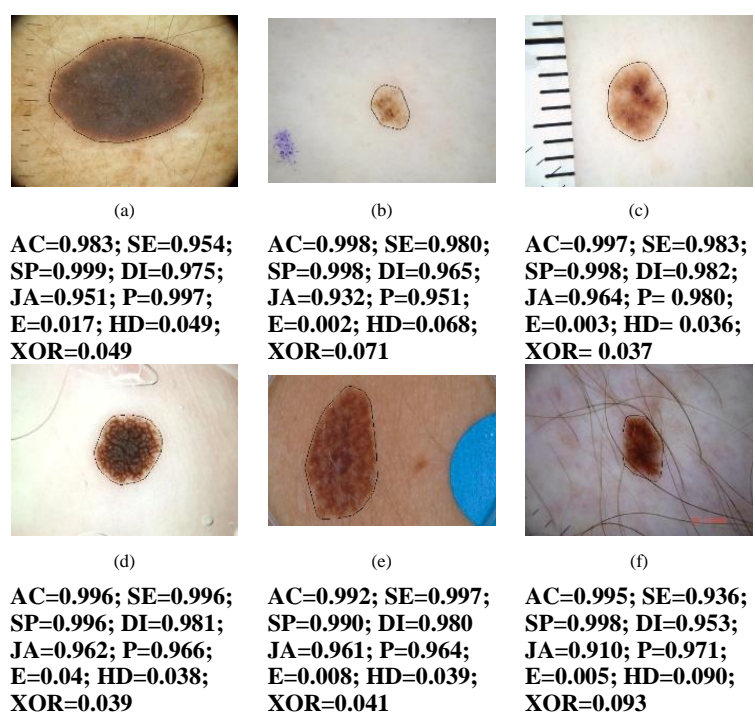

(a)
AC=0.983; SE=0.954; SP=0.999; DI=0.975; JA=0.951; P=0.997; E=0.017; HD=0.049; XOR=0.049

(b)
AC=0.998; SE=0.980; SP=0.998; DI=0.965; JA=0.932; P=0.951; E=0.002; HD=0.068; XOR=0.071

(c)
AC=0.997; SE=0.983; SP=0.998; DI=0.982; JA=0.964; P= 0.980; E=0.003; HD= 0.036; XOR= 0.037

(d)
AC=0.996; SE=0.996; SP=0.996; DI=0.981; JA=0.962; P=0.966; E=0.04; HD=0.038; XOR=0.039

(e)
AC=0.992; SE=0.997; SP=0.990; DI=0.980 JA=0.961; P=0.964; E=0.008; HD=0.039; XOR=0.041

(f)
AC=0.995; SE=0.936; SP=0.998; DI=0.953; JA=0.910; P=0.971; E=0.005; HD=0.090; XOR=0.093

**Figure 6.** Results by SCS on the image examples with artifacts and/or aberrations shown in Figure 1.

As the employed datasets consist of images of different sizes, we have selected the minimum common dimension setting *maxdim*=500. As well, given the different number of colors for the available images, a good compromise has been to set *colnum*=64. In order to determine the best perceptual parameter values, we manually minimized the resulting metrics values over the whole $PH^2$ dataset. Then, we validated those values on the ISIC2016 dataset. Since the SCS performance is slightly better when CQ and Saliency Map are computed respectively by [Achanta et al. 2009] and [Dekker 1994] and when we set the following parameters: $T_c$ =60, $T_n$ = 50, $\theta_1$ = 0.2, $T_s$=10, $\theta_2$=0.8, in the overall paper, the examples and the Tables are relative to the above-selected methods and setting.

However, SCS may fail to identify the skin lesion accurately when the pre-processing phase has not been sufficient to clean the image properly. For instance, when the hair has been only partially removed in sub-step A.2. Moreover, for a limited number of cases, the non-correct obtained result is intrinsically explained because we based the segmentation on saliency even if, for some images, saliency is not adequate to extract the desired foreground. For instance, when the highest saliency values are found in the background and saliency value within the skin lesion is extremely low. Finally, the different high levels of detail characterizing the ground truth images of the datasets (see subsection 4.1) justifies the fact that the results are sometimes very close to the ground truth images and, at the same time, sometimes the results differ more from those.

### *4.3. Quantitative evaluation*

The segmentation results obtained by SCS have been quantitatively evaluated using the images available in $PH^2$ and ISIC2016 databases [Mendonca et al. 2015, ISIC 2016], comprising 1497 images.

On these databases, we have compared the performance of SCS with that of classical saliency-based methods including MR, RBD, RC and with that of some more recent saliency-based methods specifically applied to dermoscopic images including Fan, RSSLS, Hu in terms of the following evaluation metrics: Pixel-level accuracy (AC), Sensitivity or Recall (SE), and Specificity (SP), Dice coefficient or F-measure (DI) and Jaccard index (JA) as well as Precision (P), Error (E), Hammoude distance (HD) and XOR (XOR) as defined in [Gutman et al. 2016; Hu et al. 2019] and reported in Table 1.

We remark that for the quantitative comparison, we select the above benchmark methods (MR, RDB, RC, Fan, RSSLS, Hu) among the existing saliency-methods because they show results on all images of the databases under examination ($PH^2$ and ISIC2016) and use most of the metrics we adopted. Unfortunately, it is not possible to compare the results obtained by [Olugbara et al. 2018] since the implementation of the method is not available, and the published results refer to only 120 unspecified images belonging to the two considered databases.

The validation of the proposed method has been done by applying different methods for saliency detection [Harel et al. 2007; Achanta et al. 2009] and CQ [Dekker 1994; Bruni et al. 2015, 2017; Ramella and Sanniti di Baja 2016a] as well as by using a different choice of the parameters. This validation has highlighted that SCS is essentially independent of the choice of parameters and the employed methods (for CQ and saliency computation) since it shows comparable and consistent performance with most of their possible combinations.

The average of the evaluation metrics of the considered benchmark methods and SCS for PH$^2$ and ISIC2016 are shown in Table 2 and Tables 3-4, respectively. The performance evaluation values regarding the benchmark methods have been taken from the cited papers [Yang 2013; Zhu et al. 2014; Cheng et al. 2015; Ahn et al. 2017; Fan et al. 2017; Hu et al. 2019] and [Mendonca et al. 2015]. This comparison is limited to 1100 images since the average values of the evaluation metrics are available only for PH$^2$ and the Train Set of ISIC2016. In Table 4, the average of the evaluation metrics of the considered benchmark methods and SCS for the Test Set of ISIC2016 is shown. The distribution of the average values of the quality measures (DI, HD, XOR) employed in almost all benchmarking methods is shown in Figure 7.

Note that resulting quality values for PH$^2$ in Table 2 could be improved if the convex hull is not computed in the final step since, differently from ISIC2016, the lesion's boundary of the ground truth is accurately detected. However, to compare with the results obtained on ISIC2016 images, the convex hull computation is also performed for PH$^2$ images.

## 5. Discussion and conclusions

This paper proposes the segmentation method SCS based on saliency and, therefore, is attributable to the exiting class of saliency-based methods. It should be classified as belonging to a broader class since, to the best of our knowledge, it is among the first to employ, besides the saliency, the color information that, we recall, is considered by dermatologists as essential to performance lesion detection.

SCS is an extension of a preliminary version [Ramella 2020]. It consists of two consecutive main steps: pre-processing and segmentation. The first step is devoted to image preparation. Initially, the computation burden of the segmentation process is optionally lowered by reducing the size of the images and the number of colors. Then, the image undergoes a hair removal process, after which the saliency map is computed, and the contrast of the map is increased. We note that selecting a better hair removal algorithm [Ramella 2021a] could further facilitate the obtained segmentation results since the presence of hair could firmly have a bad influence on the skin lesion region detection in the successive step. The second step is devoted to the construction of the Binary Mask, including the skin lesion. An initial Binary Mask is built to exclude dark corners and work only on the image portion, which includes the lesion and the surrounding skin through an iterated process. Successively, components not ascribed to the Binary Mask, due to their saliency level that is lower than the threshold, but with noticeable color information and not sharp color transition from foreground to background, are possibly assigned to the Binary Mask. Finally, morphological operations are applied to improve the shape of the Binary Mask foreground, select the component with a maximal area, and take its convex hull as the segmentation result. The second step (segmentation) is entirely new. It constitutes the paper's main contribution since it is based on the novel perceptual properties/criteria related to color and saliency information suitably employed according to the context. Furthermore, the segmentation process is flexible since new properties and criteria can be added, considering other types of information besides saliency and color.

We remark that SCS is independent of the adopted computation methods employed in the first step. To give evidence to this aspect, we use the proposed method as a baseline to employ different methods for color quantization and saliency computation and test the effects of this preliminary processing.

The performance of the proposed method and eventual dependencies from color and saliency computation has been extensively evaluated by considering two publicly available databases. This evaluation is mainly based on the qualitative comparison of results to the ground truth provided in the databases and quantitatively by nine commonly adopted quality measures. Finally, a different parameter setting to achieve a trade-off between quality and performance has been explored.

Based on the performance evaluation and the obtained quality values, SCS: a) turns out a robust, appropriate, and flexible method; b) achieves good quantitative results with an adequate balance. Overall, SCS has a competitive and satisfactory performance concerning other existing saliency-based methods since the probability of missing salient regions and/or detecting falsely marked salient background regions is low. Furthermore, the results have also demonstrated the effectiveness and the utility of the combined saliency and color information for skin lesion segmentation. Finally, SCS detects rapidly salient regions since the implementation does not require extensive learning (e.g., based on many parameters and labeled training images), and the execution time is rather fast.

As pointed out in Section 4.2, SCS may fail when the hair is only partially removed during the pre-processing step and/or the saliency is not adequate for the skin lesion at hand. Therefore, in future investigations, we believe there is room to improve SCS by using a more effective hair removal method, like proposed in [Ramella G. 2021a], and employing more flexible criteria to manage the special cases of saliency distribution.

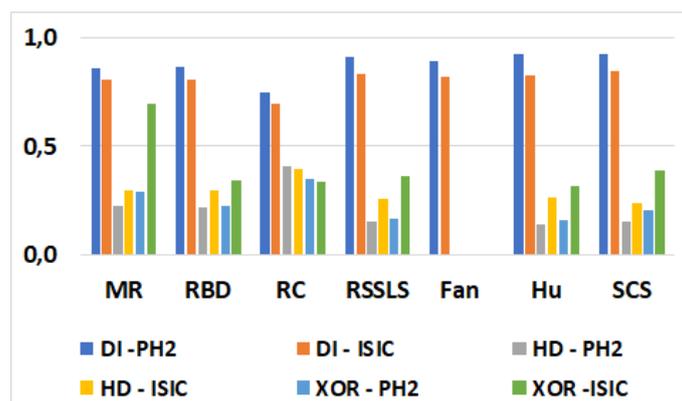

**Figure 7.** Average quality values of benchmark methods (MR, RBD, RC, RSSLS, Fan, Hu) and SCS using PH$^2$ and ISIC2016 databases.

**Table 2** – Results of PH$^2$ dataset segmentation (200 images).

| Method | AC | SE | SP | DI | JA | P | E | HD | XOR |
|---|---|---|---|---|---|---|---|---|---|
| MR [Yang et al. 2013] | - | - | - | 0.861 | - | - | - | 0.226 | 0.288 |
| RDB [Zhu et al. 2014] | - | - | - | 0.862 | - | - | - | 0.219 | 0.228 |
| RC [Cheng et al. 2015] | - | - | - | 0.748 | - | - | - | 0.410 | 0.350 |
| RSSLS [Ahn et al. 2017] | - | - | - | 0.910 | 0.682 | - | - | 0.155 | 0.165 |
| Fan [Fan et al. 2017] | - | 0.870 | - | 0.893 | - | 0.968 | 0.064 | - | - |
| Hu [Hu et al. 2019] | - | 0.942 | - | 0.922 | - | 0.913 | 0.053 | 0.139 | 0.159 |
| SCS | 0.954 | 0.952 | 0.955 | 0.921 | 0.900 | 0.921 | 0.076 | 0.155 | 0.205 |

Table 3 – Results of ISIC2016 Train Set segmentation (900 images).

| Method | AC | SE | SP | DI | JA | P | E | HD | XOR |
|---|---|---|---|---|---|---|---|---|---|
| MR [Yang et al. 2013] | - | - | - | 0.807 | - | - | - | 0.295 | 0,693 |
| RDB [Zhu et al. 2014] | - | - | - | 0.804 | - | - | - | 0.300 | 0.341 |
| RC [Cheng et al. 2015] | - | - | - | 0.697 | - | - | - | 0.393 | 0.338 |
| RSSLS [Ahn et al. 2017] | - | - | - | 0.834 | - | - | - | 0.257 | 0.362 |
| Fan [Fan et al. 2017] | - | 0.747 | - | 0.818 | - | 0.973 | 0.082 | - | - |
| HU [Hu et al. 2019] | - | 0.766 | - | 0.824 | - | 0.964 | 0.081 | 0.262 | 0.314 |
| SCS | 0.910 | 0.832 | 0.958 | 0.843 | 0.860 | 0.961 | 0.098 | 0.239 | 0.390 |

Table 4 – Results of ISIC2016 Test Set segmentation (379 images).

| Method | AC | SE | SP | DI | JA | P | E | HD | XOR |
|---|---|---|---|---|---|---|---|---|---|
| SCS | 0.962 | 0.899 | 0.975 | 0.877 | 0.891 | 0.921 | 0.045 | 0.159 | 0.303 |

**Acknowledgments**

This work has been supported by GNCS (Gruppo Nazionale di Calcolo Scientifico) of the INDAM (Istituto Nazionale di Alta Matematica).